\documentclass[aps,prl,bibliography,superscriptaddress,twocolumn]{revtex4-1}
\usepackage[utf8]{inputenc}
\usepackage{amsmath}
\usepackage{physics}
\usepackage{mathtools}
\usepackage{amsfonts}
\usepackage{amssymb}
\usepackage{braket}
\usepackage{graphicx}
\usepackage{subfigure}
\usepackage{textcomp}
\usepackage{color}
\usepackage[dvipsnames]{xcolor}
\usepackage{mathrsfs}
\usepackage{natbib}

\usepackage{makecell}

\begin{document}
\title{Characterizing Exceptional Points Using Neural Networks}
\author{Md. Afsar Reja}
\affiliation{Solid State and Structural Chemistry Unit, Indian Institute of Science, Bangalore 560012, India}
\author{Awadhesh Narayan}
\email{awadhesh@iisc.ac.in}
\affiliation{Solid State and Structural Chemistry Unit, Indian Institute of Science, Bangalore 560012, India}

\date{\today}

\begin{abstract}
One of the key features of non-Hermitian systems is the occurrence of exceptional points (EPs), spectral degeneracies where the eigenvalues and eigenvectors merge. In this work, we propose applying neural networks to characterize EPs by introducing a new feature -- \emph{summed phase rigidity} (SPR). We consider different models with varying degrees of complexity to illustrate our approach, and show how to predict EPs for two-site and four-site gain and loss models. Further, we demonstrate an accurate EP prediction in the paradigmatic Hatano-Nelson model for a variable number of sites. Remarkably, we show how SPR enables a prediction of EPs of orders completely unseen by the training data. Our method can be useful to characterize EPs in an automated manner using machine learning approaches.
\end{abstract}

\maketitle


\textit{Introduction--} In recent years, the exploration of non-Hermitian systems has been gaining wide interest~\cite{moiseyev2011non,bender2007making,ashida2020non,bergholtz2021exceptional,kawabata2019symmetry,banerjee2022non}. This is due to a large number of inherent rich physical phenomena, such as exceptional points (EPs), non-Hermitian skin effect (NHSE), non-Bloch band theory, exotic topological phases, and extended symmetry classes, which have no counterpart in the contemporary Hermitian realm. One of the most intriguing characteristics of non-Hermitian systems is the existence of EPs -- spectral degeneracy points at which not only the eigenvalues merge but also the eigenstates coalesce simultaneously~\cite{heiss2012physics,kato2013perturbation}. At EPs, the Hamiltonian matrix becomes defective. EPs are an inherent characteristic of non-Hermitian physics. They serve a pivotal role in defining non-Hermitian topological phases and the associated phase transitions. EPs can also manifest as exceptional rings~\cite{zhen2015spawning} and lines~\cite{budich2019symmetry} within the parameter space of non-Hermitian systems, and they are closely tied to PT (parity-time) symmetry-breaking phase transitions~\cite{heiss2012physics}. Non-Hermitian Bloch bands form knots and EPs play a crucial role in the transition between different knots~\cite{hu2021knots}. In addition to their fundamental significance, EPs have already found practical applications, including enhanced sensing capabilities at higher-order EPs~\cite{hodaei2017enhanced}, applications in optical microcavities~\cite{chen2017exceptional}, and the development of directional lasing technologies~\cite{peng2016chiral}. EPs have been successfully observed and studied in various experimental setups, spanning fields such as optics~\cite{miri2019exceptional,parto2020non}, photonics~\cite{ozdemir2019parity}, electric circuits~\cite{choi2018observation,stehmann2004observation}, and acoustics~\cite{shi2016accessing,zhu2018simultaneous}.

Due to its diverse applications and unique learning capabilities, machine learning (ML) is rapidly being adopted by researchers as a novel tool. In particular, ML has the potential to uncover new physics without prior human assistance. The growing trend of the application of ML techniques is not only restricted to different areas of condensed matter physics~\cite{bedolla2020machine}, but are also being explored in other areas such as particle physics, cosmology, and quantum computing~\cite{carleo2019machine}. In past years, ML techniques are being applied with outstanding accuracy to study topological phases~\cite{deng2017machine,zhuang2022classification,scheurer2020unsupervised,tibaldi2023unsupervised} and topological invariants~\cite{zhang2018machine,rodriguez2019identifying}, classification of topological phases~\cite{long2023unsupervised,scheurer2020unsupervised}, and phase transitions~\cite{carrasquilla2017machine} in the Hermitian realm. Very recently, the first applications of ML techniques to non-Hermitian systems have been undertaken. For instance, Narayan \textit{et al.}~\cite{narayan2021machine}, Cheng \textit{et al.}~\cite{cheng2021supervised} and Zhang \textit{et al.}~\cite{zhang2021machine} have undertaken the study of non-Hermitian topological phases using convolutional neural networks (NNs). Yu \textit{et al.} have used an unsupervised method, namely diffusion maps, to explore such phases~\cite{yu2021unsupervised}. Non-Hermitian topological phases in phononics have been studied using manifold clustering~\cite{long2020unsupervised}. Furthermore, Araki and co-authors have analysed NHSE in an ML framework~\cite{araki2021machine}. Moreover, ML approaches have been recently explored to investigate various non-Hermitian experimental platforms. Examples include, physics-graph-informed ML to study second-order NHSE in topoelectrical circuits~\cite{shang2022experimental}, principal component analysis and NNs to explore non-Hermitian photonics~\cite{ahmed2023machine}, and diffusion maps to analyse non-Hermitian knotted phases in solid state spin systems~\cite{yu2022experimental}.

In this work, we propose and demonstrate the characterization of EPs using NNs. We introduce a new feature, which we term \emph{summed phase rigidity} (SPR), which allows an unambiguous characterization of higher-order EPs. As an illustration of our approach, starting with simple models, we have categorised the EPs in various systems with increasing complexity with the help of NNs. In particular, we show how EPs can be distinguished in two- and four-site models by means of NNs. Furthermore, our NN construction and SPR enables an accurate prediction of the order of the EPs. Finally, using the celebrated Hatano-Nelson model for a variable number of sites, we demonstrate an accurate prediction of EPs and show how SPR allows a prediction of EPs of orders completely unseen by the training data. Our approach is useful for the characterization of EPs in an automated manner using ML techniques.



\begin{figure}[t]
\centering
\includegraphics[width=0.45\textwidth]{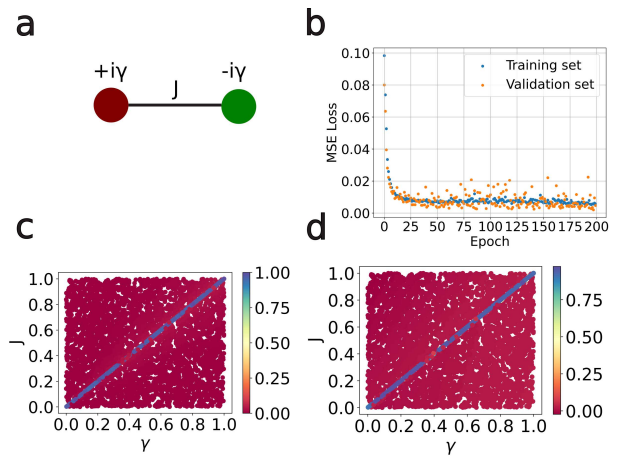}
\caption{\textbf{Training of NN for the two-site model.} (a) Illustration of the two-site gain and loss non-Hermitian model. Here $J$ is the coupling between the two sites and $\pm i\gamma$ denotes the gain and loss terms. (b) A NN with 2 hidden layers was constructed. The hidden layers consist of 16 and 4 neurons, respectively. The total number of trainable parameters is 121. The loss curve with epoch is plotted during training. The blue represents the training dataset and the yellow for validation. It is clear from the loss curve that our model is not over-fitted. (c) $-\ln(|r|)$ is plotted for the test dataset in $J-\gamma$ plane on a color scale. A color bar values tending to unity represent an EP and low values denote non-EP. (d) The corresponding predicted value from the NN is shown. We see a good agreement between the actual and predicted values in (c) and (d).}
\label{2site}
\end{figure}

\textit{Two-site model--} We first consider the simple two-site non-Hermitian model with on-site gain and loss to illustrate our approach [see Fig.~\ref{2site}(a)]. The system is described by the following Hamiltonian~\cite{zhang2020high}

\begin{equation}
    H_{2}=\begin{pmatrix}
\omega_0 + i\gamma & J  \\
J &  \omega_0 -i\gamma\\
\end{pmatrix},
\end{equation}

where $\omega_0$ is the onsite potential, $J$ is the coupling between the sites as shown in Fig.~\ref{2site}(a), and the non-Hermiticity is introduced by the gain and loss term $\pm i\gamma$. The eigenvalues are given by $\lambda_\pm=\omega_0\pm\sqrt{J^2-\gamma^2}$. For simplicity, we choose $\omega_0=0$, and the EP occurs at $\gamma=\pm J$. Note that this model can host EPs of order two only. 

We briefly describe the concept of phase rigidity, based on which we will classify the EPs. The phase rigidity, $r$, at any point in the parameter space of a given Hamiltonian is defined as~\cite{muller2008exceptional,eleuch2016clustering}

\begin{equation}
 r=\frac{\langle\Psi_{L}|\Psi_{R}\rangle}{\langle\Psi_{R}|\Psi_{R}\rangle},
\end{equation}

where $\Psi_L$ and $\Psi_R$ are left and right eigenstates of the non-Hermitian Hamiltonian.
Due to bi-orthogonality, $r$ takes a value of zero at EPs and a value of unity far from the EP. Instead of $r$, which ranges from zero to one, we propose to take the negative log of $r$, i.e., $-\ln{|r|}$. This change of scale essentially enhances the separation between EPs and non-EPs in the parameter space -- this leads to a much-improved characterization of EPs as we will show. So at an EP, this quantity takes a large positive value, and far from an EP, it drops to zero.

We choose 40,000 randomly generated points in the $\gamma-J$ plane for our two-site model and compute $-ln|r|$. Then, in a similar manner, we produced 10,000 points such that $\gamma=J$ and combined them with the points from above. Among this generated data, we kept 10$\%$ as test data, and the rest are used as training data. We constructed a network consisting of two hidden layers with 16 and 4 neurons per layer, respectively. The total number of trainable parameters is 121. We used the rectified linear unit (ReLU) as the activation function for the hidden layers. The loss (mean squared error loss) curve  for 200 epochs with a batch size of 32 is shown in Fig.~\ref{2site}(b). We note that convergence is quite fast (within nearly 50 epochs) and the loss for the test set does not fluctuate too much  or deviate from the training set loss curve, indicating the sign of a good fit.

After training is performed by optimizing various hyperparameters (see Table~\ref{table:1}), we predict the order of EPs for the test set. On the other hand, for this simple illustrative case, we already know the true order of the EPs. We compare the true and the predicted results for the same data points in the $\gamma-J$ plane, as shown in Fig.~\ref{2site}(c) and Fig.~\ref{2site}(d). We observe that our NN predictions are almost identical to the true values with an accuracy of 99$\%$, based on which we can easily classify whether a given point is an EP or not. We also calculated the performance of the network by means of the $R^2$ score. This is presented in Table~\ref{table:1} and is close to 0.968.

\begin{table}[t]
\begin{center}
\caption{\textbf{Details of NN structures.} The details of the constructed networks are presented for the different models. We note that for a fixed number (two) of layers to get a good $R^2$ score ($\approx 0.95$), one needs to increase the number of total trainable parameters by increasing the number of neurons per layer. This is because of the increasing complexity of models. Along with the $R^2$ score we have reported the mean absolute error (MAE) and the root mean square error (RMSE) for each model.}
\begin{tabular}{c c c c c c}
\hline
\hline
Model  & \thead{No. of \\hidden \\layers(neurons)} & \thead{Total\\ parameters} & \thead{$R^2$ \\score} & MAE & RMSE\\
\hline
\hline
Two-site  & 2 (16, 4)  & 121 &   0.968 & 0.018 & 0.068\\
Four-site &   2 (32, 16) & 673  & 0.974 & 0.088 & 0.181 \\
\thead{$N$ sites \\(Hatano-Nelson)} & 2 (12, 4) & 105 & 0.998 & 0.057 & 0.087\\
\hline
\hline
\label{table:1}
\end{tabular}
\end{center}
\end{table}


\textit{Four-site model--} Next, we consider a slightly more involved scenario in a four-site model, which may host both second and fourth-order EPs. The Hamiltonian reads~\cite{jaiswal2021characterizing}

\begin{equation}
H_4=\begin{pmatrix}
i\delta & p       & 0   & 0\\
p       & i\gamma & q   & 0\\
0       & q       &-i\gamma &p\\
0       & 0       & p   &-i\delta
\end{pmatrix}.
\label{eq_H4}
\end{equation}

\begin{figure}[t]
\centering
\includegraphics[width=0.45\textwidth]{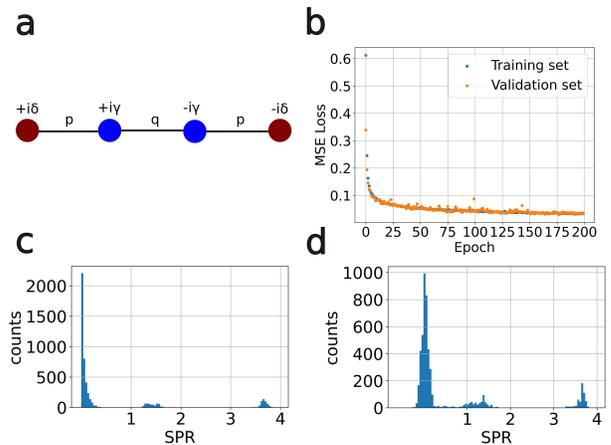}
\caption{\textbf{Training of NN for the four-site model.} 
 (a) Schematic of the four-site non-Hermitian model. Here $p$ and $q$ are the coupling constants between the sites, and $\pm i\gamma$  and $\pm i\delta$ denote the gain and loss terms. (b) An NN with 2 hidden layers was constructed. The hidden layers consist of 32 and 16 neurons, respectively. The total number of trainable parameters is 673. The loss curve with epoch during training is shown as the blue points for the training data set and as the yellow points for validation set. It is clear from the loss curve that our NN is not overfitted. (c) Histogram of test data points from the parameter space ($p,q,\delta,\gamma$) according to the SPR value. Points near SPR=0 are non-EP, i.e., ordinary points. SPR between one and two represents the second-order EPs and SPR between three and four represents the fourth-order EPs. (d) Same as (c) but now instead the predicted value of SPR from the NN is plotted. We see good agreement between actual and predicted values from (c) and (d).}
\label{4site}
\end{figure}

Here $p$ denotes the coupling between sites one and two, and between sites three and four, $q$ is the coupling between sites two and three, $\pm i\delta$ and $\pm i\gamma$ are gain and loss terms, as shown schematically in Fig.~\ref{4site} (a). The above Hamiltonian can host both second- and fourth-order EPs, depending on values of $p$, $q$, $\delta$, and $\gamma$ in the parameter space. For the rest of the discussion, we set $\gamma=1$. The EP-2 lies on a surface in the parameter space satisfying the following condition

\begin{equation}
    p^4+\delta^2+2\delta p^2-\delta^2q^2=0.
\label{eq_EP2}
\end{equation}

On the other hand, we obtain an EP-4 when both Equation~\ref{eq_EP2} and the following Equation~\ref{eq_EP4} are satisfied simultaneously,

\begin{equation}
    1+\delta^2-2p^2-q^2=0.
\label{eq_EP4}    
\end{equation}

As compared to the previous case, i.e., a classification between an EP and a non-EP, here we need to distinguish three types of points, EP-2, EP-4, and non-EPs. To do this, we have designed a new feature, which we term the \emph{summed phase rigidity}, SPR. This is defined as 

\begin{equation}
    \text{SPR}=\sum_{k} -\ln{|r_{k}|}/\text{max}(-\ln{|r_{k}|)} , 
\end{equation}

where $k$ runs over all eigenstates of the given Hamiltonian. For an $N$-th order EP, $\text{SPR}\approx N$ at the EP. The steps for constructing SPR are illustrated in Table~\ref{table:2}.

\begin{table}[t]
\begin{center}
\caption{\textbf{Constructing SPR for two site model.} To construct the SPR, we first calculate $-\ln|r|$ at the chosen points and then rescale by dividing them by the maximum value. After the rescaling, the sum of all rescaled $-\ln|r|$ gives the SPR. Note that SPR value is nearly the order of the EP at an EP and nearly zero far from the EPs.}
\label{table:2}
\begin{tabular}{c c c c c c c c}
 \hline
 \hline
 Point type &  $J$ & $\gamma$ & $-\ln|r_1|$ & $-\ln|r_2|$ & \thead{rescaled\\ $-\ln|r_1|$}  & \thead{rescaled\\ $-\ln|r_2|$} & SPR \\
\hline
\hline
Non-EP &  0.3 & 0.7 & 0.10 & 0.10 & 0.0036  & 0.0036 & 0.007\\
 
EP-2 &  0.5 & 0.5 & 27.34 & 27.34 & 0.99  & 0.99 & 1.98\\
 
EP-2 &  0.3 & 0.3 & 27.52 & 27.52 & 1  & 1 & 2\\
 
Non-EP &  0.5 & 0.8 & 0.25 & 0.25 &0.009   &.009  &0.018 \\
 \hline
 \hline
\end{tabular}
\end{center}
\end{table}

We trained the network to classify the points in the parameter space according to the SPR value. As such, we set the following cutoffs for the SPR -- an SPR value of nearly zero ($0<$ SPR $<1$) corresponds to the ordinary point (non-EP), SPR between 1 and 2 reflects a second order EP (EP-2), and SPR above three corresponds to a fourth order EP (EP-4). Note that in practical computations, SPR takes a value lower than $N$ for an $N$-th order EP. This is because our generated data points are chosen randomly and phase rigidity depends on the detuning parameters, $\delta z$, which take the system away from an EP. In general, $r\varpropto \delta z^{1/N}$. Nevertheless, we show that SPR can classify the corresponding points with an outstanding accuracy, and can thus be a very useful training feature.

We generated a data set with 50,000 points, such that it is a mixture of 5000 EP-2 and 5000 EP-4 points, with the rest being non-EPs. As before, we keep aside 10\% of data points as test data, and the rest is used as the training data. The data points are generated by randomly picking a point from the $(p,q,\delta)$  parameter space (with $\gamma=1$). The conditions for EPs are established using Eq.~\ref{eq_EP2} and Eq.~\ref{eq_EP4}. Next, we calculate SPR at these points as described above. So finally the features for each data point become $p$, $q$, $\delta$ and SPR. Our trained NN consists of two hidden layers with 32 and 16 neurons per layer respectively. The total number of trainable parameters is 673. In our training with batch size 32 and 200 epochs, the loss curve in Fig.~\ref{4site} (b) shows that the NN is not overfitted and converges rapidly. After training, we predict the SPR value for the test data set and plot it alongside the actual value in Fig.~\ref{4site}(c) and (d). We note the excellent match between the actual and predicted data. In particular, there is a clear distinction between EPs of different orders. Overall, in estimating the actual order of EPs in the test data set, we achieved an accuracy of 97.5\%.


\textit{Hatano-Nelson model with variable sites--} Next, we present the generalized approach to characterize the EPs based on the concept of SPR. We consider the paradigmatic Hatano-Nelson model~\cite{hatano1996localization}, which has served as the inception ground for many of the central ideas of non-Hermitian physics. The Hamiltonian for $L$ sites under open boundary conditions is given by the following matrix

\begin{equation}
H_{L}=\begin{pmatrix}
0   & t-\gamma/2 & 0     &...  & 0\\
 t+\gamma/2 & 0     &  t-\gamma/2 &...&  0\\
0   & t+\gamma/2   & 0     & ...&  0\\
:   & :     & :   & \ddots  &:\\
.   & .     & .   &  &.
\end{pmatrix}.
\end{equation}

Here $t\pm \gamma/2 $ denote non-reciprocal hopping terms, i.e., the unequal left and right hopping strengths. In this model, higher order EPs occur at $t=\pm\gamma/2$, and the order depends on the number of sites, i.e., $L$. Therefore, if a point is an EP in the $t-\gamma$ parameter space, then the SPR value will be nearly $L$, for an $L$ site model. For the sake of simplicity, we change the variables as $t_1=t-\gamma/2$ and $t_2=t+\gamma/2$. In this $t_1 - t_2$ parameter space, EPs occur when one of $t_1$ or $t_2$ is nearly zero and other one is nearly one.

\begin{figure}[t]
\centering
\includegraphics[width=0.45\textwidth]{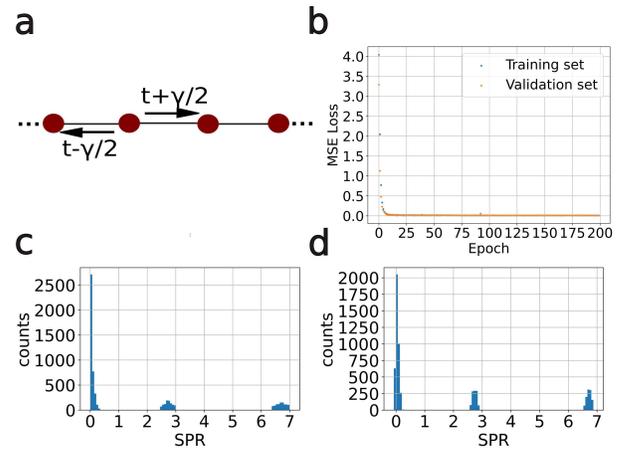}
\caption{\textbf{Training of NN for variable site Hatano-Nelson model.} (a) Schematic diagram of the Hatano-Nelson model with the left-right asymmetric hopping ($t\pm\gamma/2$). (b) A neural network with 2 hidden layers was constructed. The hidden layers consist of 12 and 4 neurons, respectively. The total number of trainable parameters is 105. The loss curve shown as a function of the epoch indicates that NN is not over-fitted or under-fitted. (c) Distribution of test data set with actual SPR values for $L=3$ and $L=7$ sites. (d) Distribution of the same data sets as (c) but with predicted SPR values from the network. We note the agreement between actual and predicted SPR values.}
\label{N_site}
\end{figure}

\begin{figure}[t]
\centering
\includegraphics[width=0.45\textwidth]{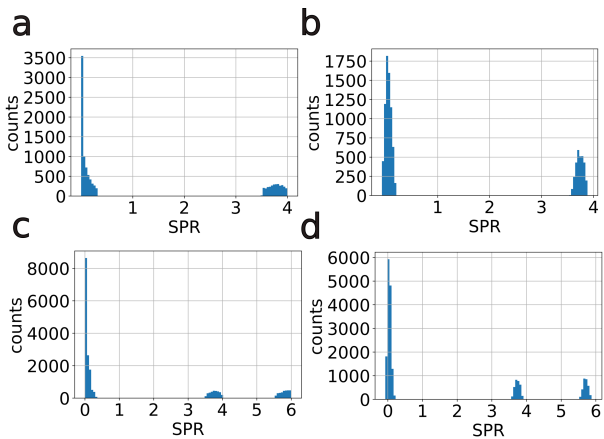}
\caption{\textbf{SPR prediction of higher-order EPs of untrained Hatano-Nelson models.} (a) Actual data distribution for $L=4$ site Hatano-Nelson model. Here, $3<$SPR$<4$ represents fourth-order EPs and SPR below one denotes ordinary points. (b) The corresponding predictions from the NN, which was trained for $L=3$ and $L=7$ sites only. (c) Actual data set for a mixture of $L=4$ and $L=6$ sites HN models, where $3<$ SPR $<4$ denotes an EP of order 4 and $5<$ SPR $<6$ represents an EP of order 6. (d) Corresponding predicted values from the NN. We discover that the NN is able to detect the order of the EP even when there is a mixture of data for EPs of different orders. Moreover, the NN is capable of predicting the EPs and their orders for those cases which were not included in either the training or the test data sets.}
\label{HN46site}
\end{figure}

Here we trained the Hatano-Nelson model with different sites. We have trained the network with $L=3$ and $L=7$. We generated the data by choosing random points in the $t_1-t_2$ parameter space and calculating the SPR. For $L=3$, we generated 30,000 points, which are a mixture of 5,000 points such that $t_1\approx 1$ and $t_2 \approx0$, 5,000 points such that $t_2 \approx 1$ and $t_1 \approx0$ and the rest are such that $t_1 \approx t_2$. Similarly, for $L=7$, 30,000 data points were generated with a similar distribution. Therefore, finally, we have 60,000 data points with different SPR values, i.e., nearly 3, 7 or 0 based on the values of $t_1$, $t_2$ and $L$. 

Among the generated data, 10\% was kept as test data and the rest is used to train the NN. We construct an NN with two hidden layers with 12 and 4 neurons, respectively. The loss curve in Fig.~\ref{N_site} (b) shows that NN is not over-fitted. After training, we predict the distribution of the test data set with respect to SPR value as shown in Fig.~\ref{N_site} (d), where $6<$ SPR $<7$ represents EPs of order seven, $2<$ SPR $<3$ denotes EPs of order three, and SPR $<1$ are non-EP points. By comparing to the actual SPR distribution of the test data set in Fig.~\ref{N_site} (c), we find good agreement between actual and predicted SPR value distribution with an accuracy close to 99.9\%. We, therefore, conclude that EPs of different orders can be successfully classified by using SPR.

Whether our NN has learned the generalised property of SPR is a crucial question. In order to understand this, we ask the already trained NN to predict the EPs and their orders in models that were not included during training. Fig.~\ref{HN46site} (a) shows the distribution of true SPR data from the $(t_1,t_2,L=4)$ parameter space for $L=4$ site Hatano-Nelson model. Here, $3<$ SPR $<4$ represents EPs of order 4, and $0<$ SPR $<1$ are non-EPs. Now we feed test this data set of $L=4$ sites to the already trained NN and the distribution of data with the predicted SPR is shown in Fig.~\ref{HN46site} (b). Remarkably, we find that the NN, which was trained for $L=3$ and $L=7$ sites is able to identify the EPs as well as their order with an accuracy greater than 99.9\%.

To further examine the robustness of our NN, we make the scenario more complicated by adding random points from $(t_1,t_2, L=6)$ parameter space to the existing $L=4$ data set. The resulting distribution of the SPR values is shown in Fig.~\ref{HN46site} (c), where $0<$ SPR $< 1$ are non-EPs, $3<$ SPR $<4$ are EPs of order 4 and $5<$ SPR $<6$ represent the EPs of order 6. The corresponding distribution for the predicted SPR value is shown in Fig.~\ref{HN46site} d. We note that our trained NN for $L=3$ and $L=7$ sites is not only capable of predicting the EPs and their orders for an untrained model, but it also has the ability to do so for a mixture of data from different non-Hermitian models with EPs of varying orders with an accuracy close to 99.9\%.
  
\textit{Training Details of NN models--} Before summarizing, we provide an overview of the technical aspects of our simulations, which were applied consistently across all NN models. We have followed a top-down approach to train the models. We started with complicated models, i.e., a large number of hyperparameters (number of layers and number of neurons in each layer) and systematically reduced the complexity until the performance declined significantly. This process resulted in the NN models which offer an optimal balance between computational cost and performance. We initiated the training process by employing the default Glorot uniform weight initializer, which randomly initializes weights from a normal distribution and sets all biases to zero. We selected the Adam optimizer for adaptive learning rate adjustment, minimizing the mean square error between actual and predicted SPR. The rectified linear unit (ReLU) activation function was utilized. Each model underwent training for 200 epochs using mini-batch training with a batch size of 32. No regularization techniques were applied. To monitor potential overfitting, separate test data were employed for all models. It is important to highlight that our machine learning models fundamentally operate as regression models, predicting SPR values. Subsequently, the specific SPR range is used to classify different orders of EPs.

\textit{Comparison with random forest regression--} To compare with more straightforward approaches we have trained a random forest (RF) regression model on the four-site model data set~\footnote[1]{We used the sklearn RandomForestRegressor class with the following hyperparameters: (i) Number of trees (\texttt{n\_estimators})= 300, (ii) the minimum number of samples required to split an internal node (\texttt{min\_samples\_split} = 20, (iii) \texttt{random\_state} = 0 to control the randomness, and (iv) The number of jobs to run in parallel (\texttt{n\_jobs}) = 3, while keeping other parameters as default.}. The data was split into 90\% for training and 10\% for testing. These hyperparameters yielded the best fit to our data, resulting in an $R^2$ score of 0.975. The performance metrics ($R^2$ score and RMSE) of the RF model is presented in Table~\ref{table:3}. The small difference in performance metrics between training and test data indicates no overfitting. The RF model achieves an accuracy of 92.5\% which is lower than the corresponding NN model.

\begin{table}[h]
\begin{center}
\caption{\textbf{Performance of the RF model.} We trained a random forest model with 300 trees. The slight difference in performance scores between the test and training data suggests that the model is not overfitting.\\}
\begin{tabular}{c c c c }
\hline
\hline
Data  & $R^2$ score  & RMSE\\
\hline
\hline
Train data & 0.975 & 0.176\\
Test data &  0.949 & 0.247\\
\hline
\hline
\label{table:3}
\end{tabular}
\end{center}
\end{table}

\textit{Discussion and summary--} In summary, we have conveyed a data-centric approach to characterize the EPs. Notably, our approach goes beyond existing machine-learning applications in non-Hermitian physics, particularly in predicting non-Hermitian topological phases~\cite{narayan2021machine,cheng2021supervised,zhang2021machine,ahmed2023machine} and NHSE~\cite{shang2022experimental,araki2021machine}. We have successfully demonstrated how ML techniques can be used to predict EPs and EP orders in various models with outstanding accuracy. For the sake of transparency, we have chosen simple yet non-trivial models with increasing complexity. For the two-site model, we trained a NN to classify EPs and non-EPs. Next, we proposed a new feature -- termed (SPR) -– and used it to distinguish between EPs of different orders. We then generalized this procedure to the celebrated Hatano-Nelson model with variable sites. Remarkably, we found that our NN models were able to predict the true order of EPs with an accuracy greater than 99\% even in cases with EPs of orders completely unseen by the training data.

Looking ahead, our work may open up interesting avenues for future explorations. Our techniques can be useful for studying the parametric dependence of EPs in higher dimensions, which can become quite intricate, especially in cases such as anisotropic EPs~\cite{ding2018experimental}. The behavior of EPs can differ based on the symmetries present in the system~\cite{mandal2021symmetry}. We envisage that generalizations of our framework may assist in characterizing nontrivial behavior in such scenarios. Additionally, EPs have intriguing connections to topological phases of non-Hermitian systems, making our method potentially useful for studying their topological properties. We hope our work motivates these promising developments.


\textit{Acknowledgments--} We have used Python~\cite{van1995python} and TensorFlow~\cite{abadi2015tensorflow} for our computations, and we are grateful to the developers. M. A. R. would like to thank Sourav Mal for useful discussions, and would also like to thank Smoky, his cat, for being a calming presence when he was writing this paper. M. A. R. is supported by a graduate fellowship of the Indian Institute of Science. A. N. acknowledges a start-up grant from the Indian Institute of Science.






\textit{Data availability--} The data and code that support the findings of this study are available at \url{https://github.com/rejaafsar/EP_ML}

\bibliography{ref_new.bib}
\end{document}